\documentclass[12pt,preprint]{aastex}

\begin{document}

\title{Temperature and Entropy Fields of Baryonic Gas in the Universe}

\author{Ping He$^{1,2}$, Long-Long Feng$^{1,3}$, and Li-Zhi Fang$^{2}$}

\altaffiltext{1}{National Astronomical Observatories, Chinese
Academy of Sciences, 20A Datun Road, Chaoyang, Beijing 100012,
China} \altaffiltext{2}{Department of Physics, University of
Arizona, Tucson, AZ 85721} \altaffiltext{3}{Purple Mountain
Observatory, Chinese Academy of Sciences, Nanjing, 210008}

\begin{abstract}

The temperature ($T$) and entropy ($S$) fields of baryonic gas, or
intergalactic medium (IGM), in the $\Lambda$CDM cosmology are
analyzed using simulation samples produced by a hybrid
cosmological hydrodynamic/$N$-body code, effective capturing
shocks and complex structures with high precision. We show that in
the nonlinear regime the dynamical similarity between the IGM and
dark matter will be broken in the presence of strong shocks in the
IGM. The heating and entropy production by the shocks breaks the
IGM into multiple phases. There are no single-value relations
between $T$ or $S$ and the mass densities of the IGM
($\rho_{igm}$) or dark matter ($\rho_{dm}$). The probability
distribution functions of the temperature and entropy fields are
long-tailed, with a power law decreasing on the sides of high
temperature or high entropy. These fields are therefore
intermittent. The mean entropy, or the cosmological entropy floor,
is found to be more than $100 \ h^{-1/3}$ keV cm$^{2}$ in all
regions when $z\leq 1$. At redshift $z \simeq 2 - 3$, high-entropy
gas ($S> 50\  h^{-1/3}$ keV cm$^{2}$) mostly resides in areas on
scales larger than $1 \ h^{-1}$ Mpc and with density
$\rho_{dm}>10^2$ (in units of $\overline{\rho}_{dm}$). Therefore,
gravitational shocks are an effective preheating mechanism of the
IGM and probably enough to provide the entropy excess of clusters
and groups if the epoch of the gas falling in cluster cores is not
earlier than $z \simeq 2 \ - \ 3$. On the other hand, at redshifts
$z \leq 4$, there is always a more than $90 \%$ volume of the low
dark matter mass density ($\rho_{dm}\leq 2$) regions filled by the
IGM with temperature less than $10^{4.5}$ K. Therefore, the
multiphased character and non-Gaussianity of the IGM field would
expain the high-temperature and high-entropy gas observed in
groups and clusters with low-temperature IGM observed by
Ly$\alpha$ forest lines and the intermittency observed by the
spikes of quasi-stellar object's absorption spectrum.

\end{abstract}

\keywords{cosmology: theory - intergalactic medium - large-scale
structure of the universe - methods: numerical}

\section{Introduction}

The standard scenario that provides the theoretical skeleton for
the formation of cosmic large-scale structures in the universe is
the hierarchical clustering scenario. This scenario assumes that
massive dark matter halos are formed via the processes of
gravitational collapse, merging of smaller structures, and
accretion. Light-emitting and -absorbing objects form when the
baryonic gas, i.e., the intergalactic medium (IGM), falls, cools,
and condenses around and within the dark halos. The evolution of
the IGM in this model is largely determined by the underlying dark
matter. It is usually assumed that the dynamical behavior of the
IGM field can be obtained from the dark matter field by a
similarity mapping between the two (e.g., Kaiser 1986). The
perturbations in the density and velocity distribution of the IGM
are considered to be the same as those of the dark matter field
point by point on scales larger than the Jeans length scales.

In the linear regime, this similarity mapping is correct. Even
when the perturbations of the initial IGM distribution are
different from those of the dark matter, linear growth modes will
lead to similarity on scales larger than the Jeans length (Bi et
al. 1992; Fang et al. 1993; Nusser 2000; Nusser \& Haehnelt 1999).
However, in the nonlinear regime, the similarity between the IGM
and dark matter fields on scales larger than the Jeans length may
not always be present. It has been shown in hydrodynamics that a
passive component generally decouples from the underlying field
during nonlinear evolution (for a review, see Shraiman \& Siggia
2000). For instance, a passive component might be highly
non-Gaussian, even when the underlying field is Gaussian
(Kraichnan 1994). This nonlinear decoupling is common in
stochastic systems consisting of a ``passive component" in an
underlying mass field.

In the system consisting of the IGM and dark matter, the IGM can
be considered a ``passive component", because the evolution of the
IGM is driven by the gravity of the underlying dark matter.
Moreover, the system of the IGM and dark matter is stochastic
because of the randomness of the initial perturbations of the
cosmic mass and velocity fields. The mass and velocity fields of
the IGM and dark matter are random variables. The importance of
the stochastic nature in the evolution of cosmic mass and velocity
fields has been emphasized in many works (e.g., Berera \& Fang
1994; Jones 1999; Buchert et al. 1999; Matarrese \& Mohayee 2002;
Coles \& Spencer 2003; Ma \& Bertschinger 2004). Therefore, in the
nonlinear regime, we expect that the similarity mapping between
the IGM and the dark matter may no longer work.

The dynamical mechanism of violating the similarity between the
IGM and dark matter has been discussed in the early study of
structure formation (Shandarin \& Zel'dovich 1989). As the dark
matter particles are collisionless, the velocity of the dark
matter particles will be multivalued at the intersection of the
dark matter particle trajectories. On the other hand the IGM,
being an ideal fluid, has a single-value velocity field. Thus, at
the intersection of the dark matter particle trajectories, the
density of gas would be discontinuous, and shocks or complex
structures occur. The development of shocks or complex structures
breaks the similarity between the IGM and dark matter and thus
leads to the statistical discrepancy between the IGM distribution
and dark matter mass field.

Strong shocks and complex structures result in the heating and the
entropy increase of the IGM. The cosmic baryonic gas involved in
strong shocks should hence undergo an entropy increase. The dark
matter remains unaffected. Consequently, the IGM will be in
thermodynamically multiple phases. The relations between the
temperature and mass density, or entropy and mass density of the
IGM will not be described by an single-value relation and will
vary from place to place. It would be interesting to study the
multiphased properties of the IGM caused by the statistical
discrepancy between the fields of the IGM and dark matter. In this
paper, we focus on the temperature and entropy fields of the IGM.

Aside from theoretical interests, this study is also motivated by
the problem of the so-called entropy excess, i.e., a high entropy
floor is observed in nearby groups and low-mass clusters (Ponman
et al. 1999; Lloyd-Davies et al. 2000). Entropy excess indicates
that some processes that violate the dynamical self-similar
scaling are important in structure formation. To explain this
result, many models for IGM preheating have been proposed, such as
photoionization heating (e.g., Efstathiou 1992; Gnedin 2000),
supernova (SN) explosions, active galactic nucleus (AGN)
activities (e.g., Cole et al. 1994; Somerville \& Primack 1999),
gravitational shocks associated with clusters (Cavaliere et al.
1998) or warm-hot gas (Valageas et al. 2003).

The decoupling of the IGM and dark matter in the nonlinear regime
inevitably leads to a dynamical break of the similarity between
dark matter and gas. Hence, the temperature and entropy fields of
the IGM formed by this evolution should be treated as the
background or cosmological ``floor''. The entropies produced by
other mechanisms are superpositions on this background. Many
previous studies assumed an uniform heating of the IGM all over
the space (e.g., Mo \& Mao 2002). However, this assumption will
not hold if the background temperature and entropy are
inhomogeneous and multiphased. Some studies of the gravitational
shock heating have assumed that the heating and entropy increase
occur only in nearby groups and clusters. This is to assume that
shock heating occurs mainly in high-density regions. However, the
decoupling evolution and gravitational shocks can happen in high
mass density regions (hosting groups and clusters), as well as in
low mass density areas. Thus, shocks can also heat the IGM at low
mass density areas.

The last motivation for this study is the progress made in the art
of numerical algorithms for hydrodynamics over the last few years.
The weighted essentially nonoscillatory (WENO) scheme (Fedkiw et
al. 2003; Shu 2003) has proved to be significantly superior over
the piecewise smooth solution algorithm containing
discontinuities. It has been successful in capturing shocks and
complex structures and in calculating the entropy in both shocked
and unshocked regions. Therefore, it would be interesting to
calculate the temperature and entropy fields of the IGM with this
code.

We analyze the temperature and entropy fields of the IGM in the
$\Lambda$CDM cosmology with simulation samples produced by a
hybrid hydrodynamic/$N$-body code. This code effectively simulates
shocks and complex structures with high precision. The paper is
arranged as follows. The basic equations and the mechanism of the
IGM-dark matter discrepancy are addressed in \S 2. In \S 3 we
present the hydrodynamic cosmological simulation scheme. In \S 4
we show the basic properties of the temperature field of the IGM.
The entropy field of the IGM is calculated and analyzed in \S 5.
Finally, in \S 6 we present our conclusions followed by a
discussion of the results.

\section{The basic equations}

\subsection{Hydrodynamic equations for the IGM}

The IGM is assumed to be an ideal fluid with polytropic index
$\gamma$. The hydrodynamic equations for the IGM in the expanding
universe can be written as
\begin{equation}
\label{hydro} \dot{U}+\partial_i f^i(U)=f(t,U)
\end{equation}
where $\partial_i\equiv\partial/\partial X^i$ ($i=1,2,3$); $X^i$
denote the proper coordinates, which are related to comoving
coordinates $x^i$ by $X^i=a(t)x^i$, $a(t)$ being the scale factor.
The quantity $U$ in equation(1) contains five components:
\begin{equation}
U=(\rho, p,{\bf v}, E)
\end{equation}
where $\rho$ is the comoving density of the IGM, ${\bf v}=\{v_i\}$
is the peculiar velocity, $E=P/(\gamma-1)+\frac{1}{2}\rho{\bf
v}^2$ is the total energy per unit comoving volume, $P=a^3p$, and
$p$ is the pressure of the IGM. The quantities $f^i(U)$ in
equation(1) are given by the conservation laws of mass, momentum,
and energy as
\begin{eqnarray}
f^1(U)&=&[\rho v^1, \rho (v^1)^2+P, \rho v^1v^2, \rho v^1v^3, v^1(E+P)]
   \nonumber \\
f^2(U)&=&[\rho v^2, \rho v^1v^2, \rho (v^2)^2+P, \rho v^2v^3, v^2(E+P)]
  \nonumber \\
f^3(U)&=&[\rho v^3, \rho v^1v^3, \rho v^2v^3, \rho (v^3)^2+P, v^3(E+P)]
\end{eqnarray}
The ``force" term $f(t, U)$ on the right-hand side of equation(1)
is given by
\begin{equation}
f(t,U)=(0,-\frac{\dot a}{a}\rho{\bf v}+\rho{\bf g}, -2\frac{\dot
a}{a}E+\rho{\bf v}\cdot{\bf g}-\Lambda_{rad}).
\end{equation}
The term of $-(\dot a/a)\rho{\bf v}$ is from the expansion of the
universe. The term of $\Lambda_{rad}$ in equation (4) is given by
the radiative heating-cooling of the baryonic gas. The
gravitational force ${\bf g}=-\nabla \phi$ is produced by the
collisionless dark matter, given by
\begin{equation}
\nabla^2 \phi = 4\pi G a^2\bar{\rho}_{tot}\delta_{tot},
\end{equation}
where the operator $\nabla$ acts on the comoving coordinate ${\bf
x}$; $\delta_{tot}=[\rho_{tot}({\bf
x},t)-\bar{\rho}_{tot}]/\bar{\rho}_{tot}$, and $\rho_{tot}$ is a
sum of the comoving baryon and dark matter mass density. The total
mean density is $\bar{\rho}_{tot}(t) =1/6\pi Gt^2 \propto a^{-3}$.
The gravitational potential $\phi$ is zero (or constant) when the
density perturbation $\delta_{tot}=0$.

\subsection{The statistical discrepancy between the IGM and dark matter}

Before embarking on the numerical calculations, we point out some
qualitative features of the IGM evolution given by equation (1).
To sketch the evolution of gravitational clustering, we ignore the
heating and cooling. Consider the case in which all thermal
processes are described by the polytropic relations $p({\bf x},t)
\propto \rho^{\gamma}({\bf x},t)$, $T \propto \rho^{\gamma-1}$, or
$T=T_0(1+\delta)^{\gamma-1}$, where $\delta =[\rho({\bf
x},t)-\bar{\rho}]/\bar{\rho}$ is the IGM mass density
perturbation. From equation (1), the momentum equation is
\begin{equation}
\frac{\partial a{\bf v}}{\partial t}+
 ({\bf v}\cdot \nabla){\bf v}=
  -\frac{\gamma k_B T}{\mu m_p} \frac{\nabla \delta}{(1+\delta)}
  - \nabla \phi
\end{equation}
where the parameter $\mu$ is the mean molecular weight of the IGM
particles and $m_p$ is the proton mass.

For growth modes in the perturbations, velocity is irrotational.
We can then define a velocity potential by
\begin{equation}
{\bf v}=- \frac {1}{a}\nabla \varphi.
\end{equation}
In the linear approximation of the temperature-dependent term,
equation (6) yields
\begin{equation}
\frac{\partial \varphi}{\partial t}-
\frac{1}{2a^2}(\nabla \varphi)^2 -
\frac{\nu}{a^2}\nabla^2 \varphi =\phi,
\end{equation}
where the coefficient $\nu$ is given by
\begin{equation}
\nu=\frac{\gamma k_BT_0}{\mu m_p (d \ln D(t)/dt)},
\end{equation}
in which $D(t)$ describes the linear growth behavior. The term
with $\nu$ in equation (8) acts like a viscosity (due to thermal
diffusion) characterized by the Jeans length $k_J^2=(a^2/t^2)(\nu
m_p/\gamma k_BT_0)$.

Since the gravitational potential $\phi$ is mainly given by the
random density field of dark matter, the term $\phi$ of equation
(8) plays the role of a random force. The nonlinear equation (8)
is exactly the stochastic force-driven Burgers equation or the
Kardar-Parisi-Zhang (KPZ) equation (Kardar et al. 1986; Berera \&
Fang 1994; Barab\'asi \& Stanley 1995). The behavior of the field
$\varphi$ given by equation (8) depends on the Reynolds number,
defined by ${\cal R}\equiv (r_c/a_0)^{4/3}$ (e.g., L\"assig 2000),
where $r_c$ is the comoving correlation length of the random field
$\phi$ and $a_0$ is the dissipation length. In the case of
equation (8), the dissipation length is given by the Jeans
smoothing equation (9), or
$a_0=\nu^{3/4}r_c^{1/2}\langle\phi^2\rangle ^{-1/4}(d\ln
D/dt)^{-1/4}$. Thus, we have
\begin{equation}
{\cal R} =(k_Jr_c)^{2/3}\left(\frac{k_J}{k}\right )^{4/3}
    \langle \delta_{dm}^2(k)\rangle^{1/3},
\end{equation}
where $\delta_{dm}(k)$ is the Fourier component of the dark matter
density contrast for the wavenumber $k$. To derive equation (10),
we assume that the gravitational potential $\phi$ is only given by
the dark matter mass perturbation.

It is well known that the Burgers equation does not reduce, but
increase initial chaos (Kraichnan 1968). When the Reynolds number
(eq. [10]) is larger than $1$, the $ \varphi $ field is in Burgers
turbulence (e.g., L\"assig 2000). That is, when the Reynolds
number is high, an initially random field always results in a
collection of shocks with a smooth and simple variation of the
field between the shocks.

The correlation length $r_c$ of the gravitational potential $\phi$
is larger than the Jeans length, i.e., $k_Jr_c >1$. Therefore,
${\cal R}$ can be larger than 1 on scales larger than the Jeans
length ($k<k_J$), even when $\delta_{dm}(k_J)$ is of the order of
1. This means that the Burgers turbulence on scales larger than
the Jeans length can occur in the IGM $\varphi$ field, when the
dark matter mass density perturbations are still quasi-linear or
weakly nonlinear ($\delta_{dm}(k) \simeq 1$) on those scales. This
results in the decoupling of IGM velocity field from that of the
dark matter on scales larger than the Jeans length. When the
Burgers turbulence develops in the IGM, the field $\varphi$ is
characterized by strong intermittency and contains many
discontinuities, or shocks. The probability distribution function
(PDF) of $\varphi$ is long-tailed (L\"assig 2000). That is, the
randomly distributed shocks and intermittent spikes lead to the
statistical discrepancy between the velocity fields of the IGM and
dark matter.

\section{Cosmological hydrodynamic simulations}

\subsection{Method of hydrodynamical simulations}

From the result of \S 2, it is clear that we need to focus on
shocks and discontinuities in our simulations. In the system of
the IGM gravitationally coupled with dark matter, the temperature
of the IGM generally is in the range $10^4-10^6$K, and the speed
of sound in the IGM is only a few km s$^{-1}$ to a few tens km
s$^{-1}$. On the other hand, the rms bulk velocity of the IGM is
of the order of hundreds km s$^{-1}$ (Zhan \& Fang 2002). Hence,
the Mach number of the gas can be as high as $\sim$100 (Ryu et al
2003). The shocks and discontinuities are extremely strong. Hence
we will not use a Lagrangian approach such as the smoothed
particle hydrodynamic (SPH) algorithm. A main challenge to the SPH
scheme is the handling of shocks or discontinuities, because the
nature of SPH is to smooth the fields considered (e.g. Borve et
al. 2001; Omang et al. 2003.)

We take a Eulerian approach instead. A well-known problem of the
Eulerian algorithm are the unphysical oscillations near a
discontinuity. An effective method to reduce the spurious
oscillations is given by designed limiters, such as the
Total-Variation Diminishing (TVD) method (Harten 1983) and
piecewise parabolic method (PPM; Collella \& Woodward 1984). A
problem of the TVD method is that the accuracy necessarily
degenerates to first order near smooth extrema (Godlewski \&
Raviart 1996). This problem will cause errors in calculating the
temperature and entropy changes because they are determined by the
difference of the thermal energy $P/(\gamma-1)$ on the two sides
of the shock. When the Mach number of gas is high, the thermal
energy $P/(\gamma-1)$ is very small compared to the kinetic energy
$\rho {\bf v}^2/2$. To overcome this problem, the essentially
nonoscillatory (ENO) and WENO algorithms were proposed (Harten et
al. 1986; Shu 1998, 2003; Fedkiw et al. 2003). They can
simultaneously provide a high-order precision for both the smooth
part of the solution and sharp shock transition(Liu et al. 1994;
Jiang \& Shu 1996).

The WENO scheme WENO has been successfully applied to hydrodynamic
problems containing shocks and complex structures, such as
shock-vortex interaction (Grasso \& Pirozzoli 2000a, 2000b),
interacting blast waves (Liang \& Chen 1999; Balsara \& Shu 2000),
Rayleigh-Taylor instability (Shi et al. 2003), and MHD (Jiang \&
Wu 1999). WENO has also been used to study astrophysical
hydrodynamics, including stellar atmospheres (Del Zanna et al.
1998), high Reynolds number compressible flows with supernova
(Zhang et al. 2003), and high Mach number astrophysical jets
(Carrillo et al. 2003). In the context of cosmological
applications, WENO has proved especially adept at handling the
Burgers equation (Shu 1999). Recently, a hybrid
hydrodynamic/$N$-body code based on the WENO scheme has been
developed. It has passed typical tests, including the Sedov blast
wave and the formation of the Zel'dovich pancake (Feng et al.
2004). This code has been successfully used to produce the
quasi-stellar object (QSO) Ly$\alpha$ transmitted flux, including
the high-resolution sample HP1700+6416 (Feng et al. 2003). The
statistical features of these samples are in good agreement with
observed features not only on second-order measures, like the
power spectrum, but also up to orders as high as 8 for the
intermittent behavior. Hence, we believe that it would be
reasonable to study the entropy and temperature fields of the IGM
with the simulation samples given by the WENO scheme.

\subsection{Samples}

For the purpose of this paper, we run the hybrid
$N$-body/hydrodynamic code to trace the cosmic evolution of the
coupled system of both dark matter and baryonic gas in a flat
low-density $\Lambda$CDM model, which is specified by the
cosmological parameters
$(\Omega_m,\Omega_{\Lambda},h,\sigma_8,\Omega_b)=(0.3,0.7,0.7,0.9,0.026)$.
The baryon fraction is fixed using the constraint from primordial
nucleosynthesis as $\Omega_b=0.0125h^{-2}$ (Walker et al. 1991).
The linear power spectrum is taken from the fitting formulae
presented by Eisenstein \& Hu (1998).

The simulations were performed in a periodic, cubic box of size 25
h$^{-1}$Mpc with a 192$^3$ grid and an equal number of dark matter
particles. The simulations start at a redshift $z=49$, and the
results are noted at the redshifts of $z=$4.0, 3.0, 2.0, 1.0, 0.5,
and 0.0. The time step is chosen by the minimum value among the
following three timescales. The first is from the Courant
condition given by
\begin{equation}
 \delta t \le \frac{ cfl \times a(t) \Delta x}{\hbox{max}(|v_x|+c_s,
|v_y|+c_s, |v_z|+c_s)}
\end{equation}
where $\Delta x$ is the cell size, $c_s$ is the local sound speed,
$v_x$, $v_y$ and $v_z$ are the local fluid velocities, and $cfl$
is the Courant number, for which we take $cfl=0.6$. The second
timescale is imposed by cosmic expansion, which requires that
$\Delta a /a <0.02$ within a single time step. The last timescale
comes from the requirement that a particle moves not more than a
fixed fraction of the cell size.

We then produced two realizations. One includes the radiative
cooling and heating, and one does not. For the latter, the initial
temperature of gas is taken to be $10^4$ K. For the former, the
atomic processes, including ionization, radiative cooling, and
heating, are modeled in the same way as in Cen (1992) in a plasma
of hydrogen and helium of primordial composition ($X=0.76$,
$Y=0.24$). The processes such as star formation, feedback due to
SNs, and AGN activities have not yet been taken into account. A
uniform UV background of ionizing photons is assumed to have a
power-law spectrum of the form $J(\nu) =J_{21}\times10^{-21}
(\nu/\nu_{HI})^{-\alpha}$erg s$^{-1}$cm$^{-2}$sr$^{-1}$Hz$^{-1}$,
where the photoionizing flux is normalized by the parameter
$J_{21}$ at the Lyman limit frequency $\nu_{HI}$ and is suddenly
switched on at $z > 10$ to heat the gas and reionize the universe.

For statistical studies, we randomly sampled 500 one-dimensional
(1D) fields from the simulation results at every redshift of z=4,
3, 2, 1, 0.5, and 0.0. Each one-dimensional sample, of size $L$=25
h$^{-1}$Mpc, contains 192 data points. For each of these points,
information about the mass density and peculiar velocity of dark
matter, and mass density, peculiar velocity, and temperature of
the IGM are recorded. All of the statistical analyses below are
based on the first realization, i.e., the one including cooling
and heating. The second one is only used to produce Figure 9,
which shows the effects of cooling and heating.

\subsection{An example of the IGM field}

We select one sample as an example of the IGM field and show in
Figure 1 the one-dimensional spatial distributions of the
temperature, mass density, and peculiar velocity of the IGM at the
two redshifts $z$=0 and $z$=0.5. For comparison, we also show the
densities and velocities of dark matter in the corresponding
panels. For identification, we use the notations $\rho_{dm}$ and
$\rho_{igm}$ for the mass densities of the dark matter and IGM,
respectively ($\rho$ in \S 2 actually is $\rho_{igm}$.) The mass
densities $\rho_{dm}$ and $\rho_{igm}$ are in units of
$\overline{\rho}_{dm}$ and $\overline{\rho}_{igm}$, which are the
mean densities of dark matter and gas, respectively. Spatial
position is described by the comoving coordinate $x$.

Figure 1 shows clearly that the relations between the density and
velocity fields of the dark matter and IGM are very complex. The
density and velocity fields of the IGM are not always faithful
tracers of the underlying dark matter fields, although in the
linear regime they are point-by-point tracers on scales larger
than the Jeans length (Bi et al. 1992; Fang et al. 1993; Bi 1993;
Nusser 2000; Matarrese \& Mohayaee 2002). Many of the
discrepancies between the IGM and dark matter fields shown in
Figure 1 can be explained by the Jeans smoothing (Theuns et al.
2000). However, Figure 1 shows that several differences between
the mass fields of the IGM and dark matter are on scales of a few
$h^{-1}$ Mpc, which is larger than the corresponding Jeans length
of the IGM.

The big jump in the IGM temperature around the position $x\simeq
13$ $h^{-1}$ Mpc is typically caused by a strong shock. The
temperature of the preshocked gas is $\sim$10$^{3}$K and increases
to $10^{6}$K by the shock. In this case, the temperature increases
by a factor of $\sim 10^3$. According to the shock theory of a
polytropic gas (Landau \& Lifshitz 1959), the Mach number should
be $M \simeq \sqrt{10^3}\simeq 30$. The value is reasonable.
Moreover, for strong shocks, the density has to increase by a
factor of $\sim (\gamma+1)/(\gamma-1)= 4$. This is also consistent
with the density field shown in Figure 1. It is also interesting
to note that the jump exists in the region of $\rho_{dm}<1$, i.e.,
not located close to filaments of dark matter. Therefore,
gravitational shocks of the IGM can develop not only around
massive dark halos but also in low-density regions.

\section{Temperature field of the IGM}

\subsection{Multiple phases in the temperature-density diagram}

In Figure 2 we show the relations between the temperature and mass
density of the IGM at $z=$4, 3, 2, 1, 0.5, and 0.0. Each panel in
Figure 2 contains 19,200 data points randomly drawn from a total
of 500 samples at each redshift. The basic feature of the
$T-\rho_{igm}$ diagram looks similar to those given by other
cosmological hydrodynamic simulations or semi-analytical models
(e.g., Ryu et al. 1993; Katz et al. 1996; Theuns et al. 1998;
Dav\'e et al. 1999; Valageas et al. 2002; Springel \& Hernquist
2002). Roughly speaking, there are two phases in the
$T-\rho_{igm}$ plane. (1) The first phase manifests a tight
correlation in the bottom left regions $T$ ($\leq10^{4.5}$K) and
$0.01<\rho_{igm}<2$, and (2) the second phase lies above the tight
correlation regions and has a scattered distribution in which the
temperature generally is greater than $10^{5}$K.

The tight correlation lines can be well approximated by a power
law as $T\propto \rho^{\alpha}_{igm}$ with $\alpha$ slightly
increasing from $\sim$0.61 at $z$=4 to $\sim$0.69 at $z$=0, which
is the same as the so-called IGM equation of state given by Hui \&
Gnedin (1997). The small discrepancy of $\alpha$ from the
adiabatic index $\gamma-1=2/3$ results from the photoionization
heating and radiative cooling (Valageas et al. 2002).

The gas in phase 2 sometimes is called the warm-hot intergalactic
medium (WHIM), as its temperature is $\sim 10^5-10^7$ K, i.e.,
higher than the gas in phase 1 by a factor of about 10 -- 10$^3$
(Cen \& Ostriker 1999). From Figure 2 we can see that the WHIM
underwent a significant evolution from redshift $z=4$ to the
present. At $z=4$ WHIM events are rare and mostly reside in the
regions of mass density, $\rho_{igm}\geq 1$, while WHIM becomes
quite common when $z\leq 2$. Note that the high-temperature
($10^6-10^7$ K) events, and hence strong shocks, can also occur in
the regions with density $\rho_{igm} \sim 1$, and even $\rho_{igm}
\simeq 0.1$. In other words, the case shown in Figure 1 is common.

Comparing with numerical results cited above by other groups,
phase 2 gas is more significant in our simulation. Especially in
the low mass density ($\rho_{igm}<1$) regions, the phase 2 in
Figure 2 is pronounced. This is because the WENO code is more
effective in capturing shocks in both high- and low-density
regions.

\subsection{Statistical decoupling between the IGM and dark matter}

From Figure 1 we have already seen the discrepancy between the IGM
and dark matter fields. The $\rho_{igm}/\rho_{dm}$ ratios are not
equal to unity everywhere. Hence, the baryonic fraction is not
always equal to the cosmological average value $f_b
=\Omega_{b}/\Omega_{m}= 0.085$. To show this deviation, we plot
the relationships between the mean baryonic fraction and
temperature in Figure 3. It is obvious that the baryon fraction at
$z \leq 1$ significantly exceeds the cosmological value $f_b$ for
the WHIM (at $10^5$K $< T < 10^7$ K). This is a well-known result
nowadays (Cen \& Ostriker 1999; Dav\'e et al. 2001). The
concentration of baryons into WHIM can be seen as follows. The gas
is compressed by a factor of $\sim 4$ by strong shocks. On the
other hand, the dark matter mass density at the pre- and postshock
is almost the same. Thus, the baryon fraction in the postshocked
IGM can be enhanced as high as $4 \times  0.085 \simeq 0.3$. This
result is consistent with what is shown in Figure 3.

However, WHIM may not necessarily reside in dense regions. In
Figure 4 we present a cumulative probability distribution
$P(>\rho_{dm})$ for WHIM ($T>10^5$ K) at different redshifts,
which is the fraction of WHIM with dark matter density larger than
$\rho_{dm}$. Note that we use the dark matter density $\rho_{dm}$
in Figure 4, not $\rho_{igm}$, as massive dark halos should be
identified by high $\rho_{dm}$, rather than by high $\rho_{igm}$.
If gravitational collapsing or collapsed regions are characterized
by $\rho_{dm}>10$, Figure 4 shows that most of WHIM at $z\leq2$
does not reside in or close to massive dark halos. In other words,
many strong shocks may not be due to the accretion by the
filaments of dark matter. One should not simply identify all WHIM
as dense clouds or massive halos.

\subsection{Probability Distribution Functions of temperature field}

The multiphased features of the temperature field can be described
by the PDF of the temperature field. The results are shown in
Figure 5. We use three density ranges: (1) low, $0.1 < \rho_{dm}
<0.4$, (2) medium, $0.4 < \rho_{dm} <2.0$, and (3) high, $2 <
\rho_{dm} < 10$. All PDFs have similar shape but shift to higher
temperature with higher density. The PDFs are sharply peaked at
some low temperatures, while some have long tails on the
high-temperature side. The tails are approximately given by a
power law $ T^{-a}$ with the index $a \simeq 4$ at $z=4$, and
$a\simeq 1.7$ at $z=0$. Therefore, the high-temperature tails are
prolonged at low redshifts. A field with such power-law-tailed PDF
is typically intermittent (Nakao 1998). We also calculate the
$T-\rho_{dm}$ diagram, which shows the similar result of
multiphased structure as the $T-\rho_{igm}$ diagram.

We show in Figure 6 the relations between the mean temperature
$\overline{T}$ and the dark matter mass density $\rho_{dm}$. One
can see that $\overline{T}$ is actually inadequate to describe the
temperature field if the field is long-tailed, or multiphased.
From Figure 6 we can see that $\overline{T}$ at redshifts $\leq 1$
and $\overline{\rho}_{dm}\sim 1$ is in the range $10^4.5-10^5$ K.
However, Figure 5 shows that in the range
$\overline{\rho}_{dm}\sim 1$ the probability for temperature of
$10^3-10^4$K is comparable to the probability for $T = 10^4-10^5$
K. Therefore, the IGM cannot be approximately modeled as a single
phase gas with temperature of either $\sim 10^4.5-10^5$ K or
$10^3-10^4$K.

In Figure 6 each panel has three $\overline{T} - \rho_{dm}$
curves, which correspond to the data smoothed by windows on scales
0.26, 1.04 and 4.17 h$^{-1}$ Mpc, respectively. It is interesting
to see that the $\overline{T}-\rho_{dm}$ relations are basically
independent of the smoothing scales. Therefore, the multiphased
feature of the IGM temperature does not come from the sampling,
but from the non-Gaussian PDF of the field. Such a field is
substantially different from a uniform or a Gaussian field.

The $\overline{T} - \rho_{dm}$ relation of WHIM given by Dav\'e et
al. (2001) is approximately $\overline{T} \propto \rho_{igm}$.
From Figure 6 we can also see $\overline{T} \propto \rho_{igm}$.
However, they are different. The mean $\overline{T}$ in Figure 6
is given by averaging over all pixels including both the WHIM and
the gas with low ($<10^5$ K) temperature, since the $\overline{T}
- \rho_{dm}$ relation for low ($<10^5$ K) temperatures is
$\overline{T} \propto \rho^{a}_{igm}$ with $a<1$. Therefore, for
WHIM only, we may have $\overline{T} \propto \rho^{a}_{igm}$ and
$a>1$. Therefore, Figure 6 shows again that our samples contain
more shock-heated events.

\section{Entropy field of the IGM}

\subsection{Entropy-temperature and entropy-mass density relations}

The entropy of the IGM is measured by the parameter
$T/n_{igm}^{2/3}$, where $n_{igm}$ is the number density of
baryons. It is actually related to the entropy density $s$ of the
IGM by $s/\rho_{igm}=\ln T/n_{igm}^{2/3} +$ const. As we will not
use $s$ below, we use this notation for the entropy parameter or
simply entropy, $S=T/n_{igm}^{2/3}$. If temperature $T$ is in
units of keV (1 keV $\simeq 1.2\times 10^7$ K), entropy $S$ is in
units of keV cm$^{2}$. Figure 7 presents the relation between the
entropy and temperature of the IGM at $z=$ 4, 3, 2, 1, 0.5, and 0
for the 19,200 data randomly drawn from the total 192$^3$ data
points.

Figure 7 also shows the two phases of the IGM: (1) a tight
correlation along a line of $S \simeq$ const at the bottom of the
figure, i.e., the entropy of the IGM is independent of IGM
temperature, and (2) a scattered distribution above the tight
correlation line. They are the same as Figure 2. The gas in phase
1 approximately satisfies $T \propto \rho_{igm}^{2/3}$. The gas in
phase 2 experiences an entropy increase process, mainly because of
the heating by shocks and complex structures in the IGM. A strong
shock increases $T$ by a factor of $10^1-10^3$, $n_{igm}$ by a
factor of $\sim$4, and therefore, the postshocked entropy can be
enhanced by a factor of $10^1-10^3$.

Phase 2 shows a significant evolution from redshift $z=4$ to the
present. There are almost no high-entropy ($S>50$ h$^{-1/3}$ keV
cm$^2$) events when $z=4$. At $z=2$ and 3, high-entropy events are
biased to $T>5 \times 10^{-3}$ keV regions. When $z \leq 1$,
high-entropy events occur in all temperatures $T>10^{-3}$ keV, or
greater than $10^4$ K ranges.

The $S$ - $\rho_{dm}$ diagram is shown in Figure 8. We again use
$\rho_{dm}$, not $\rho_{igm}$, as the clustering of the mass field
is indicated by high-$\rho_{dm}$. Although $\rho_{igm}$ does not
trace $\rho_{dm}$ point by point, one can still see the two
phases: a dense distribution along a line of constant $S$, and a
scattered distribution above the line, given by the entropy
increase because of shocks. For redshift $z<3$, the gas in phase 2
can reside in both low ($\rho_{dm}<1$) and high ($\rho_{dm}>1$)
dark matter density areas.

Figure 9a presents the mean entropy $\overline{S}$ versus density
$\rho_{dm}$. The entropy of the IGM increases by about 2 orders of
magnitude from redshift $z=4$ to 0.5. In the redshift range $3
> z> 1$, the mean entropy $\overline{S}$ is higher in higher
density areas. For dark matter mass density $\rho_{dm}>10^2$, the
mean entropy $\bar{S}$ is greater than 60 h$^{-1/3}$ keV cm$^{2}$
at the epoch $z\simeq 2$, and greater than 20 h$^{-1/3}$ keV
cm$^{2}$ at $z\simeq 3$. At $z\leq 0.5$, the shocks and
complicated structures yield an ``entropy floor'' $\overline{S}
\sim 100  - 300 \ h^{-1/3}$ keV cm$^2$ at $z\leq 0.5$ over the
entire IGM from low-density ($\rho_{dm} \simeq 10^{-2}$) areas to
high-density ($\rho_{dm} \simeq 10^{3}$) regions. This is the
cosmological background of entropy floor generated by the
nonlinear evolution of the IGM in the gravity of dark matter.

In Figure 9b we show the $\overline{S}$ - $\rho_{igm}$ relation
for samples produced by the same cosmological parameters, but
without considering radiative cooling and photoionization heating.
We see that the relation $\overline{S}$ - $\rho_{igm}$ at $z\leq
2$ in Figure 9b is about the same as that of Figure 9a. Especially
in the high-density range ($\rho_{dm}>10$), Figures 9a and 9b are
the same. Therefore, the entropy increase of the IGM at $z \leq 2$
shown in Figure 9a is not due to the heating and cooling, but
mainly determined by shocks and complicated structures. The
effects of radiative cooling and heating to the entropy evolution
are negligible at $z \leq 2$ and high-density areas. At higher
redshift $z > 2$ and low-density areas, however, the effects of
radiative cooling and heating become important.

It is interesting to compare our result of the entropy floor with
the semi-analytical estimation by Valageas et al. (2003, hereafter
VSS03). Using the data of Dav\'e et al (2001), they found that
WHIM can provide an entropy floor (or the mean $\bar {S}$) of the
order of 10$^2 \ h^{-1/3}$ keV cm$^2$ at $z\leq 0.5$, which is
about the same as Figure 9. However, their result is given by an
average of $S$ over all WHIM. As has been pointed out in \S 3.2,
most WHIM at $z \leq 2$ does not reside in high-density areas.
Therefore, it is unclear whether the high-entropy floor given by
an overall average of the WHIM is in the high-density areas.
Moreover, the average of $S$ in VSS03 is done by the method of
adiabatically bringing all patches into a same effective density.
With this method, the low-density areas would have significant
contribution to the mean $S$. In our study, the mean of $S$ is
taken in each given bin of the dark matter density, and therefore
from Figure 9 one can unambiguously conclude that the high-entropy
floor of $\sim$ 10$^2 \ h^{-1/3}$ keV cm$^2$ first formed in
high-density $\rho_{dm}>10$ areas since $z =2$, i.e., the areas
that are favored by the formation of galaxy clusters.

Moreover, the mean of $S$ given by VSS03 falls steeply with
redshift. VSS03 found $\bar{S} < 10 \ h^{-1/3}$ keV cm$^2$ at $z
\geq 1.5$. Figure 9 also shows the decrease of $\bar{S}$ with
redshift. However, it is much slower than that of the VSS03.
Figure 9 shows that the entropy floor is still $\bar{S} > 10 \
h^{-1/3}$ keV cm$^2$ when $ z=3$ and $\rho_{dm}>10$. The
difference between the redshift dependence of the VSS03 and Figure
9 is caused by two reasons. (1) The number of $\bar{S}$ of the
VSS03 is given by an average over all density regions, and Figure
9 shows that the decrease of $\bar{S}$ with redshift is faster in
lower density ($\rho_{dm}< 1$) regions; and (2) the WENO code is
more effective in describing shocks than the data used in VSS03.

\subsection{PDF of entropy field}

At first glance, the mean entropy of Figure 9 may contradict the
Ly$\alpha$ absorption in QSO's absorption spectrum. A uniform
entropy floor $S \sim $ 100 h$^{-1/3}$ keV cm$^{2}$ corresponds to
$T\simeq 1.8 \times 10^5$ K at $z=1$ and $\rho_{igm} \simeq 1$.
This temperature is obviously unacceptable with the observations
of Ly$\alpha$ forest lines. Even for a uniform entropy floor as
small as 10 $h^{-1/3}$ keV cm$^{2}$, the IGM temperature is about
4$\times 10^4$ K at $z=2$ and $\rho_{igm} \simeq 1$. This is also
contradictory with the small-scale intermittent spikes in the
transmitted flux of QSO's Ly$\alpha$ absorption. This argument is
often used to constrain preheating models. However, this argument
no longer holds if the IGM is multiphased. This can be seen from
the PDF of the entropy field.

Similar to the temperature field, the entropy field is also highly
non-Gaussian. Figure 10 presents the PDFs of the entropy field at
redshifts from $z=4$ to 0. The shapes of the PDFs at different
redshifts are similar, but the curves shift to higher $S$ at lower
$z$. The PDFs are peaked at the low-entropy end and gradually
decrease with the increasing $S$ by a power law $\propto S^{-b}$
with $b \simeq 4$ at $z \sim 4$ and $b\simeq 1.6$ at $z \sim 0$.
That is, the tail is longer when the redshift $z$ decreases.
Therefore, in terms of entropy, the IGM is also in multiple
phases. High- and low-entropy gas coexist in the IGM. In this
case, the mean entropy, $\overline{S}$, does not mean uniform
heating. For a long-tailed PDF, the mean is much larger than the
median. Thus, most of the space actually has entropy $S$ less than
the mean.

Figure 11 gives the volume filling factor $V(>S)$, which is the
volume fraction with entropy larger than a given $S$. It shows
that the volume fraction of high entropy ($S>100 h^{-1/3}$ keV
cm$^2$) is very small for all redshifts. Although the mean entropy
$\overline{S}$ is as high as $\sim 200 h^{-1/3}$ keV cm$^2$ at
$z=0$ (Figure 9a), the volume fraction $V(>200)$ is only about
10\%. Therefore, it is not surprising that the high- and
low-entropy phases can coexist in the IGM. To approximate such a
multiphased field by a uniform distribution would be misleading.

\subsection{Mass fraction of high-entropy and high-density regions}

We now study the distribution of entropy of the IGM in mass space.
We are interested in the regions with high entropy and high
density, as they provide the environment of baryonic structure
formation. Let us consider the regions with high dark matter
density $\rho_{dm}= 5$ or 10 to be the hosts of collapsed baryonic
clumps. Figure 12 shows the baryonic mass fraction $M(>S)$ of
regions with the entropy larger than a given $S$ and dark matter
density $\rho_{dm} > 5$ and $>10$, respectively.

First, we see from Figure 12 that $M(>10)\simeq M(>100)$ at $z=0$
for either $\rho_{dm} >5$ or $>10$. That is, at $z=0$, no
high-density regions $\rho_{dm} >5$ have entropy less than $100 \
h^{-1/3}$ keV cm$^2$. At higher $z$, the mass fraction with higher
entropy becomes smaller. Nevertheless, there are already $\sim
10\%$ baryons at $z\simeq 2$ to have entropy greater than $50 \
h^{-1/3}$ keV cm$^2$. This feature is not sensitive to whether
$\rho_{dm}> 5$ or 10. Since the baryon fraction in collapsed halos
is about 10\%, we may conclude that all structure formation is
probably in the entropy environment of $\overline{S} >50 \
h^{-1/3}$ keV cm$^2$ if the gas falls into the halos not earlier
than $z\simeq 2$.

In calculating Figure 12, the mass field $\rho_{dm}$ was
decomposed to the size of the resolution of the simulation, which
is 0.13 $h^{-1}$ Mpc. To study the scale dependence of the entropy
floor, we decomposed the dark matter mass field and entropy field
into different scales. This decomposition can be done by the
orthogonal scaling functions of discrete wavelet transform (DWT).
As has been showed (Xu et al. 1998), the high-density regions on
the scale 1.5 $h^{-1}$ Mpc given by the DWT decomposition can be
identified as clusters, which yields the same result as
traditional identification schemes, such as the friends-of-friends
algorithm.

Figure 13 presents the relations between the mean entropy
$\overline{S}$ and $\rho_{dm}$ for the mass field decomposed by
the DWT scaling function on scales $L=$ 0.26, 1.04 and 4.17
$h^{-1}$ Mpc. The mean $\overline{S}$ in Figure 13 is given by
averaging over all regions that are on the given scale and have
the dark matter density $\rho_{dm}$ in a given bin. Therefore,
$\overline{S}$ at $L=$ 1.04 $h^{-1}$ and $\rho_{dm}>10$ is the
mean entropy of the IGM in the regions that host the formation of
groups and clusters.

Figure 13 also shows that the $\overline{S}$-$\rho_{dm}$ relations
are basically scale-independent when $z \geq 3$. At small
redshifts $z \leq 2$, however, one can see that the larger the
size $L$, the higher the mean entropy is in high-density regions.
That is, the high-entropy IGM is biased to the places that have
high dark matter density in a large size area. This is probably
because more shocks occur around massive halos. Thus, at epochs $z
\leq 2$, massive halos are generally surrounded by high-entropy
baryonic gas. This result shows again that the entropy floor in
the regions with high mass density and on large scales is
generally larger than the estimation by an over-all average
(VSS03).

\section{Discussions and Conclusions}

We have studied the evolution of the temperature and entropy
fields of the IGM in the nonlinear regime of gravitational
clustering. We have used the $N$-body/hydrodynamic simulation
samples produced by the WENO code, which effectively describes
shocks. Strong shocks can significantly change the mass density
and velocity of the IGM fluid, at the same time having no effect
on the dark matter field. Therefore, when the shocks emerge, the
dynamical similarity between the IGM and dark matter is broken.
The gravitational shocks and complex structures of the IGM can
take place in nearby massive dark halos as well as in low mass
density regions.

The entropy production by shocks makes the IGM multiphased. The
temperature $T$ and entropy $S$ of the IGM are no longer related
to the mass density $\rho_{igm}$ or $\rho_{dm}$ by a single-value
relation. The IGM contains gas in both high- and low-entropy
phases, or both high- and low-temperature phases. For a given
$\rho_{igm}$ or $\rho_{dm}$, the IGM temperature and entropy lie
in very wide ranges from $\sim$10$^{3}$ to $\sim$10$^{7}$ K, and
from $\sim$1 to $\sim$10$^{3} \ h^{-1/3}$ keV cm$^2$,
respectively. Therefore, the temperature and entropy fields of the
IGM cannot be approximately described by a randomly uniform
Gaussian distribution.

We should point out that star formation and their feedback on the
IGM evolution are not considered in our simulation. Roughly, there
are two types of the feedbacks: (1) photoionization heating by the
UV emission of stars and AGNs, and (2) injection of hot gas and
energy by SN explosions or other sources of cosmic rays. The
photoionization heating actually can be properly considered if the
UV background is adjusted by fitting the simulation with observed
mean flux decrement of QSO's Ly$\alpha$ absorption spectrum (Feng
et al. 2003). The effect of injecting hot gas and energy by SNs is
localized in massive halos. For instance, metal absorption systems
(like CIV and MgII) in QSO spectra are generally associated with
collapsed halos (e.g., Bi \& Fang 1996). A recent Ly$\alpha$
observation around protoclusters (Adelberger et al. 2003) implies
that AGN heating does not drastically affect the gas in clusters.
Nor does it affect the IGM in low-density areas. Therefore, these
heating mechanisms are probably not strong enough to change the
basic feature of the multiphased temperature and entropy field
given in this paper.

The multiphased feature and non-Gaussianity of the IGM temperature
and entropy fields can effectively be applied to explain the
following IGM-related observations.

1. Entropy floor of groups and clusters.---For all regions on
scales greater than $1 \ h^{-1}$ Mpc and with dark matter mass
density $\rho_{dm}>10^2$, the mean entropy $\bar{S}$, or entropy
floor, is greater than 80 $h^{-1/3}$ keV cm$^{2}$ since the epoch
$z\simeq 2$ and greater than 20 $h^{-1/3}$ keV cm$^{2}$ at
$z\simeq 3$. Even for regions of $\rho_{dm}>10$, the mean entropy
still have $\bar{S} > 25 \ h^{-1/3}$ keV cm$^{2}$ at the epoch
$z\simeq 2$ and 10 $h^{-1/3}$ keV cm$^{2}$ at $z\simeq 3$ (Fig.
13). Therefore, gravitational shock is an important provider of
high-entropy gas around massive halos, which are the hosts of
clusters and groups. This result is adequate to explain the
entropy excess from the observations of clusters and groups if the
epoch of the gas falling in cluster cores is not earlier than $z
\simeq 2 \ - \ 3$ (Ponman et al. 2003). In other words, the IGM is
already significantly preheated by gravitational shocks at the
epoch $z \simeq 2 \ - \ 3$, and therefore other sources of heating
may not be important if the cluster core with entropy excess
formed after that epoch (Kay \& Bower 1999; Borgani et al 2002).

2. Ly$\alpha$ forest lines.---The Ly$\alpha$ forest lines in QSO's
absorption spectrum of $2 < z <4$ have a column density $N_{HI}$
of HI atoms in the range $13 < \ln N_{HI} <14$, which corresponds
to areas of $\rho_{dm}\simeq 1$ (Bi \& Davidsen 1997). The thermal
broadening of these lines shows that the temperature of the IGM at
this density area is about $10^4-10^5$K. The high-entropy floor
($S\sim 100 h^{-1/3}$ keV cm$^{2}$) will contradict the Ly$\alpha$
forest if the entropy field is uniform. However, the temperature
and entropy fields are highly non-Gaussian and of multiple phases.
The volume of the universe is actually dominated by regions of low
temperature and low mass density. For all redshifts $z \leq 4$,
more than 90\% volume with dark matter mass density
($\rho_{dm}\leq 2$) is occupied by the IGM with temperature less
than $10^{4.5}$K (Fig. 5). This provides the room for the QSO's
Ly$\alpha$ forests. The mean temperature is higher for lower
redshifts (Fig. 6). Therefore, the number density of the forests
lines, or absorbers, is lower at lower redshifts. These properties
match the constraints on the IGM from QSO's Ly$\alpha$ forests.
Moreover, the Ly$\alpha$ forests are very well modeled by the
semianalytical lognormal model, which assumes that the PDF of the
IGM mass density field is lognormal, and long-tailed (Bi \&
Davidsen 1997). This assumption basically is consistent with the
long-tailed PDF of IGM temperature field (Fig. 5).

3. Intermittency of Ly$\alpha$ transmitted flux.---The Ly$\alpha$
transmitted flux of the QSO's absorption spectrum is found to be
remarkably intermittent (Jamkhedkar et al. 2002, 2003). That is,
the power of the flux fluctuations concentrates in the rare
spikes. The intermittent features can be detected even on scales
corresponding to thermal broadening of gas with $T \simeq
10^3-10^4$ K. Different from Ly$\alpha$ forests, the intermittency
of the Ly$\alpha$ transmitted flux does not become weaker, but
become even stronger at lower redshifts. It is impossible to
explain the intermittency with single-phase IGM with temperature
$\sim 10^4$ K. However, strong shocks yield many big
discontinuities of the IGM temperature (Fig. 1), and therefore the
power of the temperature fluctuations is also spiky. The
long-tailed PDFs also reflect the intermittency of the temperature
field. The lower the redshifts, the longer the tails, and
therefore the intermittent features are more prominent at lower
redshifts. Moreover, the IGM at density $\rho_{dm}\simeq 1$
regions contains components of $T \simeq 10^3-10^4$ K as well as
of $T \simeq 10^4-10^5$ K. Therefore, the intermittency of the
Ly$\alpha$ transmitted flux can be well fitted with these
simulation samples (Pando et al. 2002; Feng et al. 2003).

All these results are in good agreement with the theoretical
expectation of \S 2. In the nonlinear regime, the Reynolds number
is larger and leads to the Burgers turbulence. Therefore, the
gravitational shock in the IGM provides a unified explanation of
the features of the IGM in both low- and higher density regions.
It includes the departure of the IGM-dark matter similarity
mapping, formation of the entropy excess groups and clusters, the
temperature and entropy in the regions of Ly$\alpha$ forests and
the intermittency of Ly$\alpha$ transmitted flux. The IGM-dark
matter discrepancy gives a deep understanding of the nonlinear
evolution of the system of the IGM plus dark matter.

\acknowledgments

We thank Professor C.-W Shu for helping to develop the
cosmological hydrodynamic simulation code based on the WENO
scheme. We also thank P. Jamkhedkar for help with the manuscript.
P.H. is supported by a Fellowship of the World Laboratory. L.L.F.
acknowledges support from the National Science Foundation of China
(NSFC) and the National Key Basic Research Science Foundation.
This work is partially supported by the National Natural Science
Foundation of China (10025313) and the National Key Basic Research
Science Foundation of China (NKBRSF G19990752).

\newpage

\figcaption{One-dimensional spatial distributions of temperature
$T$, mass density $\rho_{igm}$, and velocity of the IGM (solid
lines), and mass density $\rho_{dm}$ and $v_{dm}$ of dark matter
(dashed lines) at z=0 and 0.5. The densities $\rho_{dm}$ and
$\rho_{igm}$ are in units of $\overline{\rho}_{dm}$ and
$\overline{\rho}_{igm}$, which are the mean densities of dark
matter and gas, respectively.}

\figcaption{Temperature $T$ vs. IGM density $\rho_{igm}$. Each
panel is given by 19,200 points randomly drawn from the simulation
sample in 25$^3 h^{-3}$ Mpc$^3$ box with 192$^3$ data points.}

\figcaption{Relations of the mean baryon fraction with respect to
temperature of the IGM. The redshifts are taken to be 4.0, 2.0, 1.0,
and 0.}

\figcaption{Relations of the cumulated probability of WHIM
($T>10^5$ K) with respect to the dark matter density $\rho_{dm}$.
The redshifts are taken to be 4.0, 2.0, 1.0, and 0.}

\figcaption{PDFs of the IGM temperature field in the ranges of
$0.1 < \rho_{dm}<0.4$, $0.4 < \rho_{dm}< 2$, and $2 < \rho_{dm}<10$,
respectively. The redshifts are taken to be 4.0, 2.0, 1.0, and 0.}

\figcaption{Mean temperature $\overline{T}$ vs. $\rho_{dm}$ for samples
smoothed on scales $L_J=$ 0.26, 1.04, and 4.17 h$^{-1}$ Mpc, respectively.
The redshifts are taken to be 4.0, 2.0, 1.0, and 0.}

\figcaption{Entropy $S$ vs. Temperature $T$. Each panel is given
by 19,200 points randomly drawn from the simulation sample in a
25$^3 h^{-3}$ Mpc$^3$ box with 192$^3$ data points.}

\figcaption{Entropy $S$ vs. dark matter density $\rho_{dm}$. Each
panel is given by 19,200 points randomly drawn from the simulation
sample in a 25$^3 h^{-3}$ Mpc$^3$ box with 192$^3$ data points.}

\figcaption{Mean entropy $\overline{S}$ vs. $\rho_{dm}$ (a) for
the simulation with ionizing, radiation heating, and cooling at
redshifts 4.0, 3.0, 2.0, 1.0, 0.5, and 0.0, and (b) for the
simulation without ionizing, radiation heating, and cooling at
redshifts 4.0, 2.0, 0.5, and 0.0.}

\figcaption{PDFs of entropy field at redshifts 4.0, 3.0, 2.0, 1.0,
0.5, and 0.0, respectively.}

\figcaption{Volume filling factors of entropy $S$ field at
redshifts 4.0, 3.0, 2.0, 1.0, 0.5, and 0.0, respectively.}

\figcaption{Baryonic mass fraction with entropy larger than 10,
50, 100, 200, 500, and 1000 $h^{-1/3}$ keV cm$^2$, and $\rho_{dm}$
larger than 5(left panel) and 10 (right panel), respectively.}

\figcaption{Mean entropy $\overline{S}$ vs. $\rho_{dm}$ for
samples smoothed on scales $L_J=$ 0.26, 1.04, and 4.17 h$^{-1}$
Mpc. The redshifts are taken to be 4.0, 3.0, 2.0, 1.0, 0.5, and
0.}

\end{document}